\documentclass{article}

\begin{document}

\title{THE HOLE ARGUMENT. PHYSICAL EVENTS AND THE SUPERSPACE.}

\author{Mihaela IFTIME \thanks{MCPHS \& BU Center for Einstein Studies}}

\date{}

\maketitle

\begin{abstract}

The ''hole argument''(the English translation of German ''Lochbetrachtung'') was formulated by Albert Einstein in 1913  in his search for a relativistic theory of gravitation.(\cite{Einstein1913}) The hole argument was deemed to be based on a trivial error of Einstein, until 1980 when John Stachel \footnote{ Talk on "Einstein's Search for General Covariance, 1912-1915'' at the GRG meeting in Jena (1980); see \cite{Stachel1989}} recognized its highly non-trivial character. Since then the argument has been intensively discussed by many physicists and philosophers of science. ( See e.g., \cite{Brighouse1994}, \cite{EarmanNorton1987}, \cite{Earman1989}, \cite{Rovelli1999} \cite{IftimeStachel2005}, 
\cite{IftimeStachel2006}, \cite{IftimeStachel2007}, 
\cite{Rynasiewicz1992}.) 

I shall provide here a coordinate-free formulation of the argument using the language of categories and bundles, and generalize the argument for arbitrary covariant and permutable theories (see \cite{IftimeStachel2006},\cite{IftimeStachel2007}). In conclusion I shall point out a way of avoiding the hole argument, by looking at the structure of the space of solutions of Einstein's equations on a space-time manifold. This superspace $Q(M)$ is defined as the orbit space of space-time solutions on $M$ under the action of the diffeomorphisms of $M$, and it plays an important role in the study of the gravitational field and attempts to find a theory of  quantum gravity (QG).  

\end{abstract}


\pagebreak

\tableofcontents

\pagebreak

\section{Introduction. The Original Hole Argument}

The hole argument is a 'paradox', an argument against generally-covariant theories. In Einstein's most detailed account of it, $G(x)$ represents a Lorentz metric field that satisfies the field equations in the $x$-coordinate system and $G'(x')$ represents the same gravitational field in the $x'$-coordinate system; if the field equations for the metric tensor are taken to be covariant, then $G'(x)$ must also represent a  solution to these equations in the $x$-coordinate system. Do the metrics $G(x)$ and $G'(x)$ represent the same or distinct gravitational fields ?  If they represent distinct gravitational fields, as Einstein originally assumed, then the hole argument shows that no specification of the metric field outside of and on the boundary of an open subset of the space-time (a "hole") could uniquely determine the field inside the hole. 

Einstein posed the hole argument as a boundary-value problem. Hilbert(1917) had also thrown himself into the problem, and realised that it was more appropriate to formulate the hole argument as an initial value problem.\footnote{The two formulations are refered as  the boundary-value and initial-value formulations of the original hole argument. For more details see \cite{Stachel2001}}\\  

{\em The boundary-value formulation of the hole argument:}\\
Given a solution $G(x)$ everywhere outside of and on the boundary of a bounded, closed region $H$ ( the ''hole'') of space-time, including all the normal derivatives of the metric up to any finite order on that boundary, 
this data still does not determine a unique solution inside $H$ (no matter how small $H$), because an unlimited number of other solutions can be generated from it by those diffeomorphisms that are identity outside $H$ (and any number of derivatives of which also reduce to the identity on the boundary), but differ from the identity inside $H$. The resulting metric $G'(x)$ will agree with $G(x)$ outside of and on the boundary of $H$, but will differ from it inside $H$.\\

{\em The initial-value formulations of the hole argument:}\\
Assume that initially the universe was filled with matter and that later in time a hole forms. Let $G(x)$ and $G'(x)$ be two distinct solutions  of the field equations which are equal everywhere in $M$ except for a hole $H$, then if assume a spacelike (initial data) surface $S_{3}: t=0$ such that the hole is entirely in the future development $D^{+}(S_{3})$, then because the two metrics are equal everywhere outside and on the boundary of the hole, they will have the same set of initial data on the surface. \cite{Rovelli1999}

{\em The conclusion:}

The hole argument implies then that the separation between two distinct points $P$ and $Q$ both inside the hole depends on the metric, i.e., $d_{G}(P,Q)\not = d_{G'}(P,Q)$, although both metrics $G$ and $G'$ have the same initial conditions. This implies that general relativity (GR)
cannot determine the separation between two space-time points. 
No well-posed initial-value and/or boundary-value problem can be posed for Einstein's covariant equations.
Einstein found these results unacceptable and spent the next three years looking for non-generally covariant field equations. The way out of the hole argument is to conclude that two such space-time metrics $G(x)$ and $G'(x)$ must represent the same gravitational field, or in other words, that GR is a generally-covariant theory.\footnote{"General relativity" is the idea that the laws of physics are the same in all reference frames (coordinate systems). "General covariance", on the other hand, is the idea that two space-time solutions of the field equations are considered to be physically equivalent if one is isometric to the other. Some authors define "general covariance" as what we call "covariance", not making a clear distinction between the two. See \cite{Stachel1993} for this distinction.}

\subsection{Invariant Formulation of the Original Hole Argument}

Einstein's original formulation of the argument was in the language of coordinates and coordinate transformations, but here I shall use a global approach i.e. working with the whole manifold and global diffeomorphisms rather than in a coordinate patch and local coordinate transformations. \cite{IftimeStachel2006}

A general relativistic space-time is a $4$-dimensional manifold $M$, together with a pseudo-Riemannian metric field $g$ of signature $(+, +, +, -)$ on $M$. From the covariance of the field equations, if $g$ is a solution of Einstein's equations, then the pull-back (or dragged-along) metrics $\phi^{*}(g)$ obtained from $g$ by the mapping induced by diffeomorphisms $\phi:M \to M$ are also solutions of the field equations. The question at the heart of the hole argument is the following: Do all dragged-along metrics $\phi^{*}(g)$ describe the same gravitational field?  
Einstein's ultimate answer is ''yes'' ( i.e., the field equations are generally-covariant).

The hole argument is intimately related to the problem of individuation of physical events. In 1915 Einstein realized that there was a mistaken assumption about the nature of space-time and after dropping that assumption there would no longer be any incompatibility between general covariance of his field equations and determinacy. In the language of manifolds, Einstein's line of reasoning on how to avoid the hole argument translates to the fact that, at least inside the hole, the space-time points are {\em not} individuated independently of the metric field. 
Indeed, if the points of the space-time manifold $M$ are not individuated independently of the metric, it implies that when we drag-along the metric, we actually drag-along the physically individuating properties and relations of the points\footnote{This metric field represents not only the chrono-geometrical structure of space-time, but also the potentials for the inertio-gravitational field; see e.g. \cite{Stachel2000}.}. So, the pull-back metric does not differ {\em physically} from the original one. An entire equivalence class of diffeomorphically-related solutions to the field equations should corresponds to {\em one} inertio-gravitational field.

As a consequence the points of the manifold can be characterized as space-time elements, but they lack individuation as events, t.e., points of a particular space-time, unless and until the metric field $g$ is specified.
\footnote{For generic empty space-times (i.e., space-times with no symmetries) one can use the four non-vanishing invariants of the Riemann tensor to individuate the space-time points. See \cite{Stachel1993}, pp. 155-156}

\section{Further Extensions of the Hole Argument: Covariant and Permutable Theories.}

The kinematical framework of all classical fields is well described in mathematical terms as geometric structures on a fiber bundle, and their dynamics in terms of the jet extensions of the total space. One may use jet bundle formulation to construct differential operators and Lagrangians of such fields. A classical physical theory can be described mathematically by a system of differential equations on section maps of a some (gauge) natural bundle $(\mathtt{F}M\stackrel{\pi}{\longrightarrow }M)$.\cite{IftimeStachel2006} 

A natural bundle $\mathtt{F}$ is defined as a covariant functor from the category of manifolds  into the category of bundles with structure group. One can think of $\mathtt{F}$ as a rule that takes an $n$-dimensional manifold $M$ into a bundle $(\mathtt{F}M=E\stackrel{\pi}{\longrightarrow}M)$ and any (local) diffeomorphism  $\phi: M\to M$ to a bundle morphism $\mathtt{F}\phi=\Phi$ over $\phi$ and preserves composition. \footnote{ Notationally we will make no distinction between the functor $\mathtt{F}$, the fibered manifold 
$(E\stackrel{\pi}{\longrightarrow }M)$ or the total space $E$, but one should keep in mind that $F$ is a functor, i.e. the object $(E\stackrel{\pi}{\longrightarrow }M)$ together with a class of automorphisms.} From the definition, a natural morphism $\Phi:E\to E$ is the (natural) lift of a local base diffeomorphism $\phi$, and they are the only bundle morphisms that preserve the natural structure on the total space $E$.\cite{KMS1993}
A gauge-natural bundle can be similarly defined as a covariant functor from the category of principal bundles and 
principal fibre-preserving morphisms into the category of fibred manifolds and fibre-preserving morphisms. Examples of gauge-natural bundles are principal bundles\footnote{The linear frame bundle $LM\to M$, which is the principal bundle associated to the tangent bundle $TM$, is actually a natural bundle)\cite{KMS1993}, \cite{Fatibene}}. A natural bundle $(E\stackrel{\pi}{\longrightarrow}M)$ is completely defined by its base manifold $M$, i.e., the local trivializations and the bundle transition functions can be canonically constructed out of the atlas of local charts on $M$. \footnote{On the other hand, there are a whole family of natural bundles defined on the same base manifold $M$, e.g. the tangent and cotangent bundles, and all tensor bundles.} 

Natural bundles provide a modern description of the classical picture of geometric objects in differential geometry and general relativity ( e.g. linear connections on manifolds, metric fields), while gauge-natural bundles represent the arena for gauge objects ( e.g. principal connections) that appear in gauge field theories. At the level of physics, a {\em type of theory} is correlated with the concept of a type of (gauge-) natural bundle. We can think of the total manifold $E$ as the set of all possible choices of the physical field-values at all points of base manifold $M$ and a given section $\sigma$ as a particular choice of a " physical field" ( of the type of theory represented by the natural bundle) over $M$.\footnote{Typically we will consider $M$ to be a $4$-dimensional space-time manifold $M$ ( but could be any $n$ dimensional manifold).} A section $\sigma$ of a natural bundle $F$ is also called a geometric (or natural) object ( e.g, tensor fields, derivative operators, which satisfy a quasi-linear\footnote{linear in first derivatives} field equation \cite{Geroch}. A {\em mathematical model} of a type of physical theory corresponds to a (global) cross-section of the natural bundle, or a class of gauge-equivalent cross sections of the gauge-natural bundle what describes the corresponding a type of physical theory. A {\em particular physical theory of given type} is represented by a rule\footnote{In order to formulate the rule of selecting cross-sections one needs to use jet extensions of the configuration space \cite{Sardanashvily}, but we do not need go into any further details about the rule for selecting cross-sections.} for selecting a class of cross-sections (models) of that type of theory. 

{\em Example 1:}  Maxwell's electromagnetic theory can be described as a
rule for selecting the class of cross-sections of the gauge-natural bundle of one-form fields that obey the linear, gauge-invariant field equations derived from the Maxwell
lagrangian. \\
Born-Infeld electrodynamic theory is a rule for selecting a (different) class of cross-sections of the same gauge natural bundle of one-form fields that obey the
non-linear, gauge-invariant field equations derived from the Born-Infeld lagrangian \cite{BornInfeld}. 

{\em Example 2}: GR is a generally-covariant theory and it can be formulated in terms of natural bundles. There are two main ways to formulate GR.

One mathematical model for the gravitational field uses only the metric and the second order prolongations ( see e.g., \cite{HE}) and it is given in terms of $G$-structrures or the corresponding holonomy groups. One advantage of using $G$-structures to model the space-time metric structure is that the gravitation can be represented as a gauge field (like any other classical fields \cite{Sardanashvily}).  In this context, a gravitational field $g_{ij}$ is defined as a cross-section $\sigma$ of the associated bundlle $(LM/O(1,3)\to M)$ of orbits under the action of orthogonal group $O(1,3)$ on the linear frame $LM$, with the standard fiber $GL(4)/O(1,3)$. \\
An $O(1,3)$- structure on $M$ given by such a global cross-section 
$\sigma$ specifies an equivalence class of $O(1,3)$-related linear frames at each point of $M$. Locally, $\sigma$ can be represented by a family of local cross-sections $\sigma_{U_{i}}: U_{i}\to LU_{i}$; two such local cross-sections $\sigma_{U_{i}}$ and $\sigma_{U_{j}}$ being related on the overlapping open set $U_{i}\bigcup U_{j}$ by a local gauge transformation: $\sigma_{U_{i}}(p)=\sigma_{U_{j}}(p)\circ \rho_{ji}(p)$, where $\rho_{ij}(p)\in O(1,3)$, for any $p\in U_{i}\bigcup U_{j}$.

Another mathematical model of the gravitational field uses the metric, and the connection, and it can be constructed on a fibered bundle $(E\to M)$, over the $4$-manifold of space-time events $M$. The total space $E$ is a $54$-dimensional space\footnote{The fiber over $p\in M$ consists of pairs $(g, \nabla)$ where $g$ is pseudo-Riemannian metric and $\nabla$ a derivative operator at $p\in M$. The dimension of the fibres in this case is $50=(10 + 40)$. A (torsion-free) derivative operator at a point in $M$ could be defined, for example, as a map from covector fields on $M$ to second-rank covariant tensors at $p$, subject
to additivity, the Leibnitz rule, and consistency with the exterior derivative.}. A cross-section $\sigma = (g, \nabla)$ of $E$ is a map $\sigma: M\to E$ such that $\pi\circ \sigma$ is the identity map on $M$.
A given cross-section $\sigma = (g, \nabla)$ represents a
particular choice of the " gravitational field" (of the type represented by the bundle) over $M$ if it satisfies the following first order quasi-linear equations: \cite{Geroch}

\begin{equation}
\nabla_{a}g_{ab} =0
\end{equation}

\begin{equation}
R^{e}_{ab(c}g_{d)e} =0
\end{equation}

\begin{equation}
R_{c(ab)}^c = 0
\end{equation}

\subsection{Other Covariant Theories.}

If $\mathcal T$ is a theory on a manifold $M$ defined  as a rule for selecting a set of models of a type of theory for a natural bundle$(\mathtt{F}M=E\stackrel{\pi}\to M)$, then for any arbitrary diffeomorphism of $\phi: M\to M$ there is a uniquely defined fibered manifold automorphism 
$\Phi=\mathtt{F}\phi: E\to E$ that projects over $\phi$. For any local cross-section $\sigma$ of $\pi$ defined on a small open set $U_p$ around $p\in M$, then the pulled-back cross-section
$\phi^{*}\sigma =\Phi\circ\sigma\circ\phi^{-1}$ is a new local cross-section of $\pi$
defined on the open set $U_{q}=\phi(U_{p})$ around $q=\phi(p)\in M$.
A theory $\mathcal T$ is covariant if all the pull-back cross-sections $\phi^{*}\sigma$ of any model $\sigma$ in the theory are also models of the theory. A covariant theory is generally-covariant if all the pull-back cross-sections of any model $\sigma$ in the theory represents the same physical model. (See \cite{IftimeStachel2006} for details.)

\subsection{Permutable Theories.}

The hole argument that applies to fibered manifolds can be modified to apply to fibered sets; the concept of a covariant theory resulting in the concept of a permutable theory. 

In this case, a theory $\mathcal T$ over a set of elements $M$ is a given by a rule for selecting a class of cross-sections of a fibered set $(E\stackrel{\pi}\to M)$, where $\pi$ is a surjective map. In analogy to the geometric (natural) object case, we restrict ourselves to the case in which to each base automorphism there corresponds a lift to a 'natural' fiber automorphism of the fibered set.

The theory $\mathcal T$ is called {\em permutable}  if whenever a cross-section belongs to this class, so does  every cross-section that results from applying a fiber automorphism to it. This results in a class of automorphically-related cross-sections. $\mathcal T$ is a {\em generally-permutable} theory if all the  members of this class are semantically identical, and so represents the same physical model.

Let $M=\{a_{1},a_{2},..., a_{n}\}$ be a set of elements of the same kind, and let us denote with $F$ the set of states, properties-values, or  processes of an arbitrary element in $M$ \footnote{ this space generalizes the phase and path space (in classical mechanics) and the Hilbert (in quantum theory)}. In the case, $E$ is taken to be the Cartesian product $M\times F$ and the projection map $\pi$, the projection $pr_1$ over $M$. A {\em mathematical model} of such a type of theory $(E\to M)$ is represented by a  cross-section $\sigma(a_{i})=(a_{i}, f_{i})$ of $E$, which represents a selection of a state, process, property-value $f_i$ at an arbitrary point $a_{i}\in M$.

A very important case (for physics) is to consider a class of fibered objects in which the base is a set
$S \neq \emptyset$ abstracted from its topological and differential structures,  while a fiber over a point $X_a$ is a differentiable manifold. The total space is $\displaystyle X= \coprod_{a\in S}X_a$ and a cross-section $\sigma : S \mapsto X$ takes each point $a\in S$ into an element of the fiber over $a$. This case has important applications to the quantum mechanics of many-particle systems, and in particular, to the cases in which these particles are all of the same kind, and it discussed in more detail in \cite{IftimeStachel2007}.

\subsection {Diffeomorphisms: Passive and Active Diffeomorphisms. Background-Independent theories.}

General covariance is a key feature of the Einstein field equations. This implies that they take the same form in any coordinate system or, expressed in geometrical language, that they are invariant under the group of $4$-dimensional diffeomorphisms. A diffeomorphism $\phi\in Diff(M)$ is represented locally by smooth maps $x^{i}\longmapsto y^{i} = \phi^{i}(x)$.

Physicists use two interpretations of the concept of
diffeomorphism-invariance: ''passive'', and ''active''. ''Passive'' diffeomorphism invariance refers to invariance under a change of coordinates, i.e., 
$x \longmapsto y$. ''Active'' diffeomorphism-invariance on the other hand relates different objects in $M$ in the same coordinate system, i.e. interpreting $\phi$ as a map associating two different points on the manifold $M$.

Any theory can be made invariant under ''passive'' diffeomorphisms. 
(because coordinates do not have a physical meaning).  

A background independent theory is a physical theory defined on a base manifold $M$ endowed with no extra structure, such as geometry or fixed coordinates. If a theory does include any such geometric structures, it is called background-dependent. Theories like QED, QCD are theories on a fixed (flat or curved) background space-time metric. GR or in any general relativistic theory on the other hand are distinguished from other dynamical field theories by invariance under ''active'' diffeomorphisms; its field equations are invariant under all differentiable diffeomorphisms ( the group $Diff(M)$) of the underlying manifold $M$, which have no spatio-temporal significance until the dynamical fields are specified. 

In a background-independent theory, there is no kinematics independent of the dynamics. Any background independent theory is generally covariant. For background independent theories (e.g., GR), the gauge freedom is the full diffeomorphism group $Diff(M)$ of the base manifold $M$ and two mathematical models$(M,\sigma)$ and $(\phi(M)= M, \phi^{*}\sigma)$ are physically indistinguishable. If the theory is background dependent then the hole argument fails automatically. (see \cite{IftimeStachel2006} for details)

The quantization of the gravitational field is often described as one of the greatest challenge to theoretical physics of our time. The search for a quantum theory of gravity meets nowadays with an explosion of different types of quantizations techniques.  There are many attempts to quantize the gravity, but there is no fully accepted QG.\footnote{As Ambjorn, Juriewicz and Loll ( 2005) state "there is not a single theory of QG that is both reasonable complete and internally consistent mathematically (see \cite{Loll} p.5 )} One important difficulty that QG meets today is that such a futue theory should be background independent ( generally covariant).\footnote{''A conceptually complete theory of QG cannot be based on a background dependent perturbation theory .... In ... a complete formulation the notion of string-like particles would arise only as an approximation, as would the whole notion of classical space-time''(Michael Green 1999), \cite{Green}}

Also related to general covariance of Einstein's equations some conceptual and technical difficulties arise when one studies the equations using purely mathematical methods or computer calculations.  This is mainly due to the fact that there is no preferred way of splitting the $4$-dimensional space-time into time and $3$-dimensional space, and the problem of choice of a good time coordinate. While it is possible to 'slice up' space-time into a family of $3$-dimensional slices, everyone is free to choose their own slicing of space-time, and geometrically natural choice of slicing can help towards an elegant and useful description of solutions of Einstein's equations. 

Also related to the Cauchy problem for the Einstein equations: the initial data on a space-like hypersurface are subject to four constraint equations, which must be solved in the course of finding a pair of ''true observables,''freely specifiable as ''positions'' and ''velocities'' initially, the evolution of which off the initial hypersurface should be uniquely determined by a pair of coupled, nonlinear field equations. This program has been carried out with locally-defined variables.
only in some highly idealized situations( e.g., cylindrical waves), in general, quantities expressing the degrees of freedom and the equations governing their evolution are highly non-local and can only be specified implicity (e.g., in terms of the conformal 2-structure coordinate and velocity; see \cite{dinverno})

\section{Physics and the Geometry of Superspace}

Superspace is the space of all possible space-time geometries on a 4-dimensional differentiable manifold.( see e.g.,\cite{Fisher}, 
\cite{IsenbergMarsden}) This space plays an important role in the study of the gravitational field at the classical and quantum levels. Superspace can be constructed as an orbit space; such that one point in the superspace is an equivalence class of diffeomorphically-related (isometric) space-time metrics. The construction gives a way of identifying inequivalent space-time metrics with different gravitational fields. In other words, physics is the geometry of the superspace - the set of all distinguisable physical states. 

In more detail, denote by $\mathcal{M}(M)$ the collection of all pseudo-Riemannian metrics on a space-time manifold $M$ and $Diff(M)$ the group of diffeomorphisms of $M$ \footnote{$Diff(M)$ is a smooth manifold modelled on the space $\mathcal{X}(M)$ of vector fields on $M$}. The group $Diff(M)$ acts as a transformation group on $\mathcal{M}(M)$ by pulling-back metrics on $M$: for all $\phi\in Diff(M)$ and $g\in \mathcal{M}(M)$ the action map is defined by $(\phi, g)\longmapsto \phi^{*}(g) $. \footnote{$Diff(M)$ acts naturally on all tensor bundles over $M$ by differentiation, and so on cross-sections of these bundles.}

For a fixed metric $g$, $\mathcal{O}_{g}=\{\phi^{*}(g)|\quad \phi\in Diff(M)\}$ is the orbit through $g$. Two metrics $g_1$ and $g_2$ are on the same orbit, if and only if $\phi^{*}(g_1)=g_2$ for some  $\phi\in Diff(M)$, i.e. $g_1$ and $g_2$ are isometric metrics.

If $g_1$ satisfies Einstein's equations  then $g_2$ does. Two space-time solutions of Einstein's equations are considered to be {\em physically equivalent} if one is isometric to the other. 

The action of $Diff(M)$ on $\mathcal{M}(M)$ partitions $\mathcal{M}(M)$ into (disjoint) isometry classes of space-time metrics.

The superspace is the space $\mathcal{Q}(M)=\mathcal{M}(M)/Diff(M)$ of all isometry classes of space-time solutions on $M$ i.e., the set of physically distinguishable states. A physical space-time represents a point in $\mathcal{Q}(M)$. 
The projection map $\Pi: \mathcal{M}(M)\longrightarrow \mathcal{Q}(M)$ identifies all isometric pseudo-Riemannian metrics to a single diffeomorphically-equivalence class. The image of $\Pi(g)=[g]$ represents the (unique) physical gravitational field defined by $g$.
\cite{Fisher}, \cite{IsenbergMarsden}

{\em The Moduli Space of Cross-sections.}

Let us consider a type of theory defined on a (gauge-) natural bundle $(E\stackrel{\pi}\to M)$, and let $\mathcal T$ be a background independent\footnote{The hole argument does not apply to background-dependent theories} theory on $M$ i.e., a rule for selecting a set of cross-sections of $(E\stackrel{\pi}\to M)$.

The concept of superspace can be generalised to the case of space of cross-sections of $(E\stackrel{\pi}\to M)$. If we denote $\Gamma(E;M)$ the collection of all cross-sections of $\pi$, the diffeomorphisms group $Diff(M)$ acts as a transformation group on $\Gamma(E;M)$ by
pulling-back\footnote{See e.g., \cite{KMS1993} for a distinction between pulling-back forms and pushing-forward vectors.} cross-sections of $E$: for all $\phi\in Diff(M)$ and $\sigma\in \Gamma(E;M)$ the action map is defined by $(\phi, \sigma)\longmapsto \phi^{*}(\sigma) $. 

For a fixed model $\sigma$, $\mathcal{O}_{\sigma}=\{\phi^{*}(\sigma)|\quad \phi\in Diff(M)\}$ is the orbit through $\sigma$. Two cross-sections $\sigma_1$ and $\sigma_2$ are on the same orbit, if and only if $\phi^{*}(\sigma_1)=\sigma_2$ for some  $\phi\in Diff(M)$. The action of $Diff(M)$ on $\Gamma(E;M)$ partition $\Gamma(E;M)$ into (disjoint) diffeomorphically equivalent classes of cross-sections.

The space $\Gamma(M)$ of all diffeomorphically-equivalent classes of cross-sections on $M$ is $\Gamma(E;M)/Diff(M)$. 
The projection map $\Pi:\Gamma(E;M)\longrightarrow \mathcal{Q}(M)$ identifies all diffeomorphically-equivalent models to a single diffeomorphically-equivalence class. The image of $\Pi(\sigma)=[\sigma]$ represents the physical model defined by $\sigma$.

To analyse the structure of quotient spaces such as superspace one may use the concept of a "slice"\footnote{The slice for the action of a group $G$ on a manifold $M$ at a point $p\in M$ is a transversal manifold $S_p$ to the orbit at $p$.}, which may not exist in general case \footnote{See \cite{Palais}, \cite{IsenbergMarsden} for the proof of the existence of slice for the action of diffeomorphisms group in 3-dimensional Riemannian case and some modified superspace of the 4-dimensional space of solutions of Einstein's vacuum field equations.} When there is slice, then one may decompose (stratify) the superspace
into manifolds of diffeomorphically-related cross-sections ( e.g., space-time solutions). The existence of such a stratification is usually shown by proving the existence of slices at every point for the group action. \footnote{For details about the slice theorem for the superspace of Riemannian metrics, see \cite{Palais}}
Though it has a topological structure,\footnote{The strongest topology in which the projection map $\Pi:\Gamma(E;M)\longrightarrow \mathcal{Q}(M)$ is continuous} the superspace is not a manifold and the infinite-dimensional analysis is quite complicated.

\section{Some Open Problems}

An open problem related to the general covariance of the Einstein equations is to understand the implications of the space-time metric on the topology of the underlying manifold. If $g_1$ and $g_2$ are two space-time metrics satisfying the same Einstein's equations, they are considered physically equivalent if one is isometric with the other, i.e., there is a global diffeomorphism $\phi:M\to M$ such that $g_{2}=\phi^{*}g_{1}$ . Implicitly, the topology of the underlying space $M$ is fixed. But different topologies arise for different solutions of Einstein's equations, even if the metrics are locally diffeomorphic.
Given a fixed space-time 4-manifold $M$ the question is how many Einstein's non-isometric solutions 'live' on $M$?
This problem is related to the stability problem of space-time solutions, which is one of the most important unsolved problems of GR. Example of stable solutions are the Minkowski space-time(see \cite{Klainerman}) and Schwarzschild black hole(see \cite{Vishveshwara}, but in general this is a very complicated problem.

Another open problem is the study of the structure of the space of space-time solutions. In practice in solving the Einstein equation for example, one does not start from a prescribed global manifold $M$, rather one solves the equations locally, and then looks for the maximal extension of this solution subject to some criteria for this extension, e.g., null and/or timelike
geodesic completeness \cite{MacCallum}. This situation is somehow similar to the ability of gluing morphisms and objects, similar to the local affine spaces into schemes, and maybe can be 
formulated in terms of stacks of moduli spaces of cross-sections. The idea is to analyze the geometry of the superspace in terms of stacks, that is to collect together the individual superspaces defined on fixed topological spaces.

\section{Acknowledgments}

I would like to thank the conference organisers for their kind invitation to attend the conference and to present the paper, and Dr. John Stachel for telling me about the hole argument and for his collaboration and numerous suggestions.


\begin{thebibliography}{100}

\bibitem{BornInfeld}\textbf{Kerner, R., Barbosa, A.L., Gal'tsov, D.V. } (2005)
\textit{Topics in Born-Infeld Electrodynamics}, hep-th/0108026

\bibitem{Brighouse1994}\textbf{Brighouse, C. } (1994),
\textit{space-time and holes,} in D. Hull, M. Forbes and R. M. Burian (eds.) PSA 1994 Vol.{\bf 1} pp. 117-125.
 

\bibitem{EarmanNorton1987}\textbf{Earman, J. and Norton, J.} (1987), 
\textit{What Prince space-time Substantivalism? 
The Hole Story}, British Journal for the Philosophy of Science {\textbf 38}, pp. 515-525

\bibitem{Earman1989} \textbf {Earman, N.} (1989) {\em Coordinates and Covariance: Einstein's view of space-time and the modern view},  Foundations of Physics, {\bf 19}, 1215-1263

\bibitem{Einstein1913}\textbf{Einstein, A.}(1914), {\em Die formale Grundlage der allgemeinen Relativittstheorie.} Kniglich Preussische Akademie der Wissenschaften (Berlin), Sitzungsberichte: pp. 1030-1085.


\bibitem{Einstein1952}\textbf{Einstein, A.} (1952),
\textit{Reltivity, the Special and General Theory}, ${\bf 15} ^{th}$ edition; translation by Robert W. Lawson, University of Sheffield; Crown Publishers Inc., New York (1961)

\bibitem{Fatibene} \textbf{Fatibene, L., Francaviglia, M.}(2004)
\textit{Natural and Gauge Natural Formalism for Classical Field Theories: A Geometric Perspective Including Spinors and Gauge Theories} Kluwer Acad. Publ.

\bibitem{Fisher}\textbf{Fisher, A.} (1969)
\textit{The Theory of Superspace},Proceedings of the Relativity Conference in the Midwest Cincinnati, Ohio, pp. 303- 357

\bibitem{Geroch}\textbf{Geroch, R.}(2004)
\textit{Gauge, Diffeomorphisms, Initial-Value Formulation, Etc.} in 
The Einstein Equations and the Large Scale Behavior of Gravitational Fields: 50 Years of the Cauchy Problem in General Relativity. Edited by Piotr T. Chrusciel and Helmut Friedrich, Birkhäuser Verlag, Basel, Switzerland, p.441

\bibitem{Green}\textbf{Green, M.}(1999)
\textit{Superstrings, M-theory and quantum gravity}, Classical and Quantum Gravity, {\bf 16}, A77-A100

\bibitem{HE} \textbf{Hawking, S.W., Ellis, G.} (1973),
\textit{The Large Scale Structure of the Universe} Cambridge Univ. Press

\bibitem{IftimeStachel2006}\textbf{Iftime, M, Stachel, J.} (2006),
\textit{ The Hole Argument for Covariant Theories} GRG, Springer, 
vol.{\bf 38}, no.{\bf 8}, gr-qc/0512021


\bibitem{IftimeStachel2007}\textbf{Iftime, M, Stachel, J.} (2007),{\em Permutable Theories and Quantum Particles} to be submitted to GRG, Springer


\bibitem{Klainerman}\textbf{Klainerman, S. , Christodoulou, D.} (1993), 
\textit{The Global Nonlinear Stability of the Minkowski Space}, Princeton Univ. Press


\bibitem{KMS1993}\textbf{Kol\'{a}\~{r}, I., Michor, P. and Slov\'{a}k, J.} (1993), \textit{Natural  Operations in Differential Geometry}, Berlin et al: Springer-Verlag


\bibitem{dinverno} \textbf{D'Inverno, Stachel, J.} (1978), 
\textit{Conformal Two Structure as the Gravitational Degrees of Freedom in GR} Journal of Mathematical Physics {\bf 19}: 2447-2460.

\bibitem{IsenbergMarsden}\textbf{Isenberg, J., Marsden, J.E.},(1982), 
\textit{A slice theorem for the space of solutions of Einstein's equations},Physics Report, 89,
{\bf no. 2}, 179-222, North-Holland Publishing Company

\bibitem{KMS1993}\textbf{Kol\'{a}\~{r}, I., Michor, P. and Slov\'{a}k, J.} (1993), 
\textit{Natural  Operations in Differential Geometry}, Berlin et al: Springer-Verlag

\bibitem{Lewandowski}\textbf{Kami{\'n}ski, W., Lewandowski, J., Bobie{\'n}ski,M.}(2005)
\textit{Background independent quantizations: the scalar field I}, gr-qc/0508091

\bibitem{Loll}(2006)\textbf{ Ambjorn,J., Jurkiewicz J., Loll, R.}
\textit{The Universe from Scratch,} Contemp. Phys. {\bf 47}, 103-117


\bibitem{MacCallum}\textbf {Stephani, H. et al},(2003),
\textit{Exact Solution of Einstein's Field Equations}, ${\bf 2}^{nd}$ Cambridge Univ. Press


\bibitem{MacLane1992}\textbf {MacLane, S.},(1992),
\textit{Sheaves in Geometry and Logic: A first introduction to topos theory}, New York: Springer-Verlag

\bibitem{Palais}\textbf{Palais, R.}, (1970)
\textit{Notes on the slice theorem for the space of Riemannian metrics} Letter circulated in 1969 and notes written at Santa Cruz in 1975

\bibitem{Rynasiewicz1992}\textbf{Rynasiewicz, R. }(1992)
{\em Rings, Holes and Substantivalism: On the Program of Leibniz Algebras,} Philosophy of Science, {\bf 45}, pp. 572-589

\bibitem{Rovelli1999}\textbf{Gaul, M, Rovelli, C.}(1999)
\textit{Loop Quantum Gravity and the Meaning of Diffeomorphism Invariance}
Authors: Marcus Gaul, Carlo Rovelli
Lect.Notes Phys. {\bf 541}, (2000) 277-324

\bibitem{Sardanashvily}\textbf{Sardanashvily, G. and Zakharov, O.} (1991)
\textit{Gauge Gravitation Theory}, World Scientific


\bibitem{Stachel1986}\textbf{Stachel, J.} (1986), 
\textit{What a Physicist Can Learn From the Discovery of General Relativity}
in Remo Ruffini (ed), \textit{Proceedings of the Fourth Marcel Grossmann Meeting on General Relativity}
Amsterdam: Elsevier Science Publishers, 1857-1862.  


\bibitem{Stachel1989}\textbf{Stachel, J.}(1989)
\textit{Einstein's Search for General Covariance, 1912-1915} 
in Don Howard and John Stachel, eds. Einstein and the History of General Relativity, Einstein Studies, vol. {\bf 1}, Boston/Basel/Berlin: Birkhäuser, pp. 63-100; 
reprinted in Stachel, John, \textit{Einstein from 'B' to 'Z'} , 
Boston/Basel/Berlin: Birkhäuser 2002, pp. 301-337


\bibitem{Stachel1993}\textbf{Stachel, J.} (1993)
\textit{The Meaning of General Covariance: The Hole Story} 
in John Earman, et al., eds., Philosophical Problems of the Internal and External Worlds. 
Pittsburgh: University of Pittsburgh Press/Konstanz: Universitätsverlag, pp.129-160.


\bibitem{Stachel1999}\textbf{Stachel, J.}(1999)
\textit{New Light on the Einstein-Hilbert Priority Question}
Journal of Astrophysics and Astronomy {\bf 20} pp. 91-101;\\
(reprinted in \cite{Stachel2002}, pp. 353-364

\bibitem{Stachel2000}\textbf{Stachel, J.}(2000)
\textit{The Story of Newstein}, 
Genesis of General Relativity Sources and Interpretation, vol {\bf 4}, \textit{Gravitation in the Twilight of Classical Physics: The promised Mathematics}Dordrecht/Boston/London, Kluwer Academic Publishers, (Berlin, Springer 2006)


\bibitem{Stachel2001}
\textbf{Stachel, J.} (2001)
\textit{The Cauchy Problem in General Relativity: The early years}, pp.405-416 in \textit{Proceedings of the Seconf International Conference on the History of General Relativity}, eds J. Eisenstaedt and A. Kox. Boston/Basel;/Stuttgart: Birkhäuser. To appear in \textit{Going Critical} ( in press)

\bibitem{Stachel2002}\textbf{Stachel, J.}
\textit{Einstein from 'B' to 'Z'}, (Boston/Basel/Berlin: Birkhäuser 2002)

\bibitem{Stachel2005}\textbf{Stachel, J.}(2005), 
\textit{Structure, individuality and Quantum Gravity}
to appear in French, gr-qc/0507078

\bibitem{IftimeStachel2005}\textbf{Stachel, J., Iftime, M} (2005),
\textit{ Fibered Manifolds, Natural Bundles, Structured Sets, G-Sets and all that: The Hole Story from Space Time to Elementary Particles,} gr-qc/0505138

\bibitem{Stachel2002}\textbf{Stachel, J.}(2006)
\textit{The first two acts}, vol {\bf 1}, pp. 81-111
Renn, J., Schimmel, M.eds
\textit{The Genesis of General Relativity}, vol. {\bf 3},
\textit{Gravitation in the Twilight of Classical Physics: Between Mechanics, Field Theory and Astronomy,} \\
reprinted in \textit{Einstein from 'B' to 'Z'}

\bibitem{Vishveshwara} \textbf{Vishveshwara, C.} (1970)
Phys. Rev. D1: 2870 


\end{thebibliography}
\end{document}